\documentclass[12pt]{article}
\usepackage{latexsym}
\usepackage{amsmath}
\usepackage[dvips]{graphicx}
\usepackage{textcomp}

\newcommand{\be}{\begin{equation}}
\newcommand{\ee}{\end{equation}}
\newcommand{\ba}{\begin{array}}
\newcommand{\ea}{\end{array}}

\author{Fabio Cardone $^{1,2,5}$, Giovanni Cherubini $^{3,4}$,
Andrea Petrucci $^{1,5,*}$\\ \\ $^{1}$Istituto per lo Studio dei
Materiali Nanostrutturati (ISMN — CNR) \\ Via dei Taurini - 00185
Roma, Italy
\\ $^{2}$GNFM, Istituto Nazionale di Alta Matematica "F.Severi" \\
Citt\`a Universitaria, P.le A.Moro 2 - 00185 Roma, Italy \\
$^{3}$ARPA Radiation Laboratories Via Montezebio - 01100 Viterbo,
Italy \\ $^{4}$ Facolt\`a di Medicina, Universit\`a degli Studi "La
Sapienza" \\ P.le A. Moro, 2 - 00185 Roma, Italy \\
$^{5}$Dipartimento di Fisica "E.Amaldi", Universit\`a degli Studi
"Roma Tre" \\ Via della Vasca Navale, 84 - 00146 Roma, Italy \\ *
Correspondent author: petruccia@fis.uniroma3.it}
\date{}
\title{Piezonuclear Neutrons}
\begin{document}
\maketitle \abstract{We report the results of neutron measurements
carried out during the application of ultrasounds to a solution
containing only stable elements like Iron and Chlorine, without any
other radioactive source of any kind. These measurements, carried
out by CR39 detectors and a Boron Triflouride electronic detector,
evidenced the emission of neutron pulses. These pulses stand well
above the electronic noise and the background of the laboratory
where the measurements were carried out.}

\section{Introduction}
The application of ultrasounds of suitable frequency and amplitude
to a liquid, with gas dispersed in it, brings about the process that
is known as cavitation~\cite{cg}\cite{bren}. It occurs when the
micro bubbles dispersed in the liquid collapse under the spherically
symmetric compressions of ultrasounds. The processes that go on
during the collapse and the collapse itself are quite complex and
there is a good deal of research going on in order to clarify their
physical and chemical aspects. Some of the studies that have been
carried on, have the target to exploit cavitation as the mean to
induce deuterium-deuterium nuclear fusion in a liquid
matrix~\cite{tale1}-\cite{tale5}. It is known as well that
mechanical waves like ultrasounds and shockwaves can induce or
better catalyse nuclear reaction in radioactive elements, fission in
fissile elements with emission of ionising and neutron radiation too
low for the occurring transmutations or with no emission at
all~\cite{uru1}-\cite{uru3}. The research that we have carried on,
although it might seem to deal with the same physical terms such as
nuclear reactions, nuclear radiation and to point towards the same
technological direction, it is based on different theoretical
concepts, that are all presented
in~\cite{carmig1}\cite{carmig2}\cite{carmig3}, and in this sense is
moving along a parallel path with respect to the other research
paths and together with them it contributes to enlarge our view and
knowledge of these new physical phenomena. We carried out five
experiments in the last few years. In the first three of them
~\cite{carmig4}\cite{carmig5}\cite{carmig6} we collected evidences
of anomalous production of intermediate and high mass number
nuclides within samples of cavitated water. These outcomes, that
agree with those obtained by Russian teams~\cite{uru1}-\cite{uru3},
point out that ultrasounds can induce nuclear phenomena such as
modifications of the nuclei, and alter secular equilibriums. A
further outcome of these three experiments is that the number of
protons after cavitation  is conserved while the number of neutrons
is not. This circumstance convinced us to carry out some experiments
in order to confirm this evidence by revealing the presence of
emitted neutrons during cavitation. Two sets of
experiments~\cite{carmig3} were carried out in which we cavitated
water and different solutions of metallic salts, of different
concentrations with diverse ultrasonic power and different geometry
of the sonotrode tip and the cavitation chamber. All of these
experiments succeeded in detecting neutrons. We would like to stress
that all our experimental equipment and our measurements, when
devoted to prove neutron emission, never involved any radioactive
source or unstable nuclide unlike other
experiments~\cite{tale1}-\cite{tale5}.

\section{Initial evidence of neutron emission}
\begin{figure}
\begin{center} \
\includegraphics[width=0.8\textwidth]{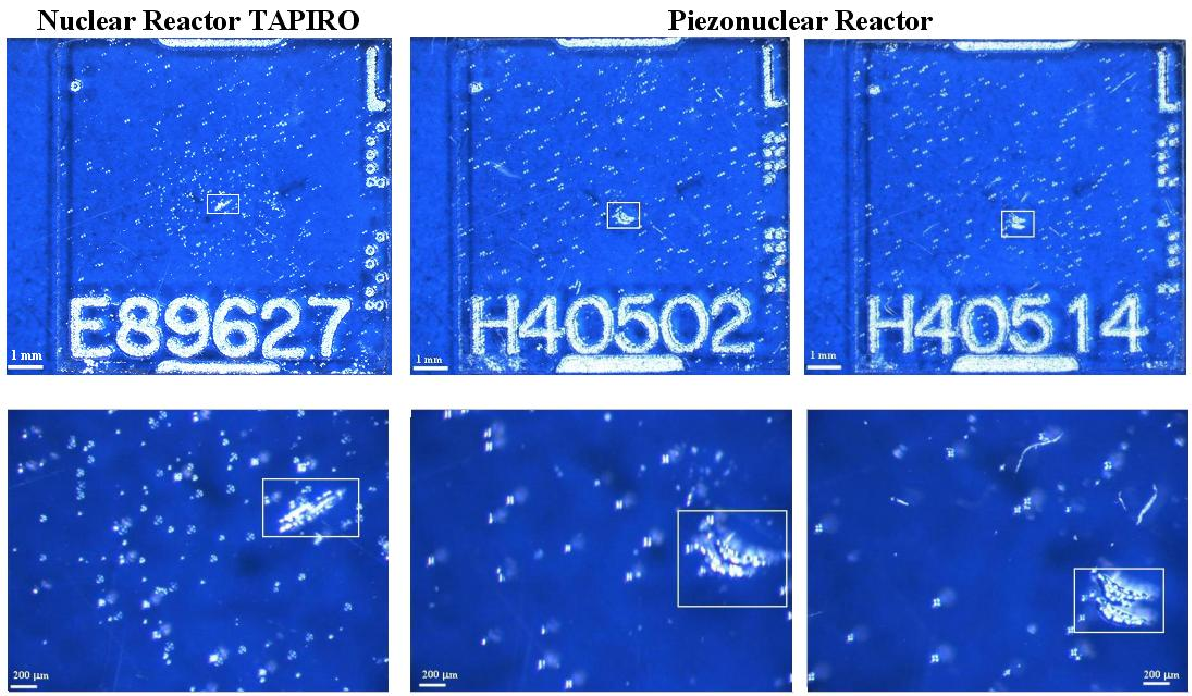} \caption{The first column shows the Boron-CR39 detector irradiated by
 the nuclear reactor TAPIRO at Casaccia ENEA with a neutron dose of 2.1 $\mu$Sv (upper row: magnification X 10
  - lower row: magnification X 50), the second and third columns show the two Boron-CR39 detectors that were
  next to the cavitation chamber during the application of ultrasounds (upper row: magnification X 10 - lower row:
  magnification X 50).}\label{CR39}
\end{center}
\end{figure}

Our first main goal was only to reveal (not exactly measure) any
possible emission of neutrons from the solutions subjected to
cavitation. Thus we used the CR-39 (PADC) plastic track detector
that is a C$_{12}$H$_{18}$O$_7$ polymer with density 1.3 g/cm$^3$
which is used for registration of heavy charged particles and is a
very convenient mean of detection. Charged particles are registered
directly, and neutrons are detected through a secondary recoil
particles or nuclear reactions. The CR39 energy range sensitivity is
very wide, from tens of KeV to hundreds of MeV. Particle tracks on
the detector become visible after chemical etching  and are
investigated using a microscope. As we stated above, the only
evidence that we could gain from the previous experiments was the
non conservation of the number of neutrons which suggests a possible
neutron emission, but does not say anything about their spectrum,
their isotropy and homogeneity in space and their constancy in time.
In this sense, a passive detector like the CR39, which is able to
integrate the signals in a wide range of energy regardless of the
time structure of the emission, is very useful to reveal the
presence of these apparently emitted neutrons as it does not require
any sort of adaptive electronic calibration which would be necessary
with an electronic detector in order to track and follow an emission
that, as far as we know, could be the most variable one in terms of
energy and time. Once that an initial but solid evidence of neutron
emission was gathered by these passive detectors, we soon increased
the quality of our investigation by moving to electronic Boron
Trifluoride detectors whose evidence will be presented in this
paper. In order to detect neutrons by the CR39 we used the nuclear
reaction $^{10}B$(n,$\alpha$)$^7Li$ and hence spread a 2 mm layer of
natural Boron (80.1\% $^{11}B$, 19.9\% $^{10}B$) on the CR39
detecting surface which had to convert neutrons into alpha
particles, following a well known
technology~\cite{tommasino,khan,izer}. These Boron-CR39 detectors
were used in our experiment during the cavitation of a 10 ppm
solution of Iron Chloride FeCl$_{3}$ in which we conveyed ultrasonic
power for 90 minutes. The ultrasounds released in the cavitation
chamber had a stable power of 130 Watts at a frequency of 20 KHz in
order to induce cavitation according to the Rayleigh-Plesset
equation~\cite{bren}. Two Boron-CR39 detectors were placed facing
each other one at each end of a diameter of the cavitation chamber
and both facing the area were cavitation took place at 2 cm from it.
Of course two more CR39 detectors without Boron layer were placed
beside each of them in order to compare the response. Two more
detectors, one with an unscreened CR39 and the other with Boron-CR39
were used as reference and placed in a different area with respect
to the zone where cavitation was taking place. In order to have an
idea of what the traces should look like on these detectors after
etching, four more detectors were irradiated by neutrons using as
source, the fast neutron nuclear reactor TAPIRO at Casaccia ENEA
Rome, the neutron equivalent dose conveyed onto the detectors was
2.1 $\mu$Sv through a diagnostic neutron channel\footnote{Not
knowing what kind of neutron spectrum to expect from the cavitated
solution, as already stated, we decided to produce our comparison
model of traces by a source whose spectrum were the widest possible,
i.e. a nuclear reactor. According to reference~\cite{tommasino},
these kind of detectors can detect fast, epithermal and thermal
neutrons with different sensitivities of course. Hence the integral
effect, due to almost the whole neutron spectrum on the detectors,
would be traces whose quantity and shape would be compared to those
obtained from the piezonuclear reactor.}. In Fig.\ref{CR39} we
present the outcomes of these series of experiments. The Boron-CR39
with code E89627 was placed with its detecting surface orthogonal to
the neutron channel of the nuclear reactor TAPIRO with its centre on
the channel axis. It clearly shows the expected tracks at its centre
(a 50 magnification photo is presented as well for a better vision).
The other three CR39 used to measure the background around the
nuclear reactor do not show any thick track and the number of tracks
on them is much lower than that on the E89627. The CR39 detectors
with code H40502 and H40514 were used with the cavitation chamber
and show tracks absolutely compatible with that produced by the
neutrons of the nuclear reactor\footnote{The chemical etching of the
CR-39 foils was carried out in a 6.25N NaOH (Carlo Erba standard)
solution at 90$^\circ$C for 4.5 hours according to the
specifications given by the firm FGM Ambiente which provided them}.

We would like to highlight one difference between the E89627 and the
other two CR39 (H40502 and H40514). It is quite reasonable to state
that the distribution of the tracks on the central part of the
E89627 is nearly circular around the thick track. This is the
consequence of the cylindrical channel through which it was
irradiated, that produced a neutron flux with a cylindrical
symmetry\footnote{The axis of the cylinder was perpendicular to the
CR39 squared chip and passing through its centre.}. Hence, this
distribution and the thick track were gradually generated along the
irradiation time. Conversely, the other two CR39 do not show any
particular distribution of tracks which is consistent with the lack
of any preferred direction of neutron emission from the volume where
cavitation was taking place. As to what one might expect, the
emission should be isotropic. Despite that, on the two CR39
detectors there are two thick tracks at the centre of the chips
which perpendicularly faced the centre of the cavitation volume. We
reckon that this difference can be considered as a strong hint to
state that the neutron emission is not continuous but, conversely,
takes place with neutron pulses.

\section{Measurements of the emission of neutron pulses}

\begin{figure}
\begin{center} \
\includegraphics[width=0.8\textwidth]{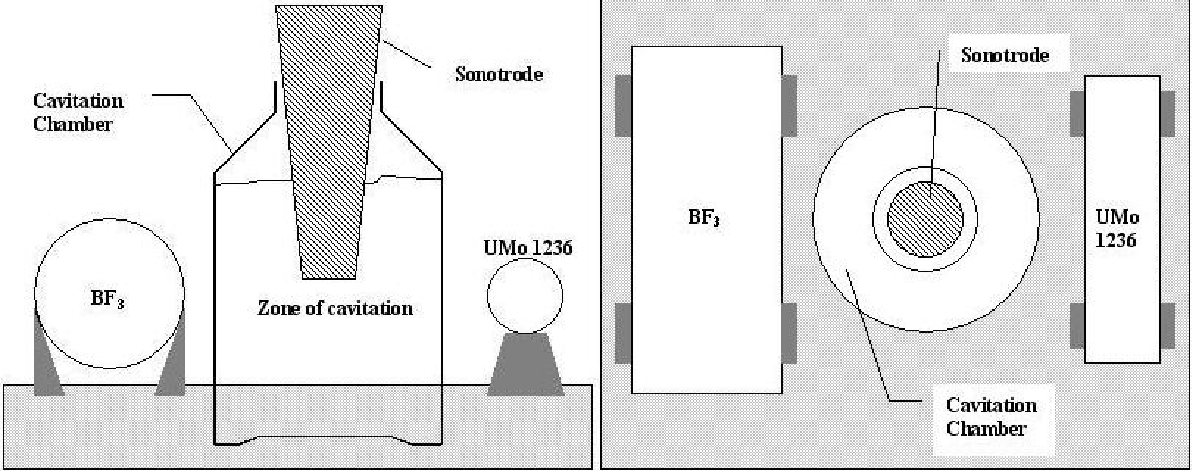} \caption{Vertical and horizontal
lay-out of the experimental set-up}\label{layout}
\end{center}
\end{figure}

\begin{figure}
\begin{center} \
\includegraphics[width=0.8\textwidth]{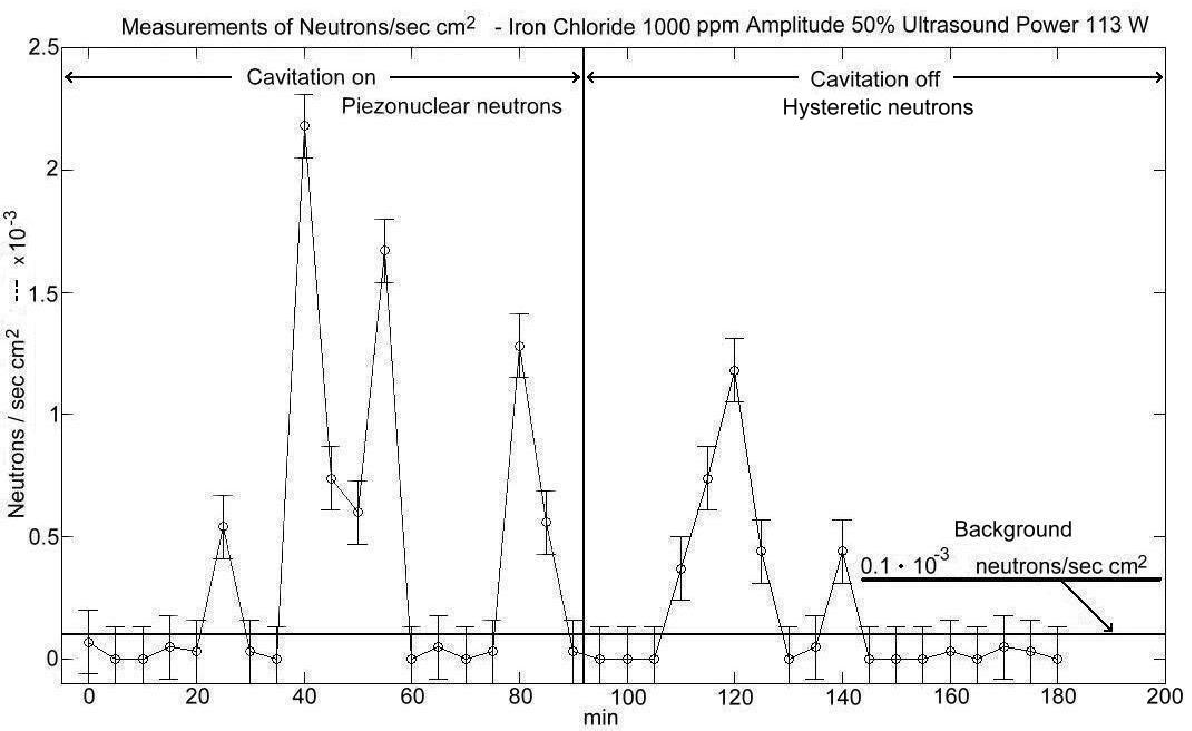}
\caption{The graph shows the neutron pulses obtained during one of
the cavitation runs. Time in minutes is on the x-axis and neutron
flux (neutrons/s cm$^2$) $\cdot$ 10$^{-3}$ is on the y-axis. The
error bars represent the sum of the pessimistic measured electronic
noise of the whole measuring equipment and the pessimistic measured
laboratory background flux i.e. 0.13 $\cdot$ 10$^{-3}$ counts/(sec
cm$^2$).}\label{graph}
\end{center}
\end{figure}

Having achieved several positive evidences of neutron emission from
cavitation, we decided to refine our research by performing the same
measurements by a Boron Triflouride detector. Before presenting the
experiments and their results, we would like to stress once more
that the purpose of these experiments has been to measure the
emission of neutrons during the cavitation of solutions of stable
elements. It might seem, on a experimental basis, that our work made
its first move from that stream of research which investigates the
induction of nuclear fission in fissile elements and/or nuclear
fusion in light elements~\cite{tale1}-\cite{uru3} by mechanical
compressions like ultrasounds. However, before presenting our
experimental outcomes, we would like to warn the reader that, both
on the theoretical and experimental side, our research diverges from
that stream and it concentrates on absolutely stable
elements\footnote{These stable elements are all but deuterium, since
we are not looking for Deuterium-Deuterium fusion.} according to the
suggestions of our theoretical
predictions~\cite{carmig1}-\cite{carmig3} which indicate that
nuclear processes can be in induced in stable nuclides too only by
suitable mechanical waves, i.e. without any application of
radioactive or nuclear active substances. We cavitated 250 ml of a
solution of Iron Chloride FeCl$_3$ by applying to it 130 stable
Watts of ultrasonic mechanical vibration at 20 KHz for 90 minutes.
The cavitation chamber was a vessel of Schott Duran\textregistered
$\ $BoroSilicate glass~\cite{duran} and the truncated conical
sonotrode that conveyed ultrasounds was made of AISI grade 304
steel. The immersion of the sonotrode in the solution was 5 cm
\footnote{Further details about the geometry of the sonotrode and
the cavitation chamber and about their exercise are contained in
three patents owned by the Consiglio Nazionale delle Ricerche (CNR)
(National Council of Researches of Italy) now published
in~\cite{patent1,patent2,patent3}.}. During cavitation we measured
neutron emission by the Wedholm Medical 2222A Boron Trifluoride
neutron monitor~\cite{wedholm} and ionising radiation and in
particular gamma radiation by UMo LB 1236 monitor whose energy range
is from 30 KeV up to 2 MeV and which gives both equivalent dose and
equivalent dose rate in a wide range from 50 nSv/h up to 10
mSv/h~\cite{berthold}. The neutron monitor was calibrated by an
Americium Beryllium standard source contained within a suitable lead
and steel shielding box in order to obtain outside of it and in
contact with it an equivalent dose rate of neutrons of 1.5 $\pm$ 0.2
$\mu$Sv/h \footnote{This equivalent dose rate value is given by the
manufacturer and refers to a new source. The Americium-241 Beryllium
neutron source, that was used for the BF$_3$ calibration, was a four
year old standard source which, considered that the Americium-241
half life is of 432.2 years, can be considered as new.}. The BF$_3$
neutron monitor was placed in contact with this shielding box that
contained the AmBe source and its position was decided in order to
present to the neutron flux a geometrical effective surface of 142
cm$^2$ perpendicular to it. We registered the number of counts from
the BF$_3$ within 100 seconds. Many of these counting runs were
carried out (sample size $\geq$ 30) and we found out that the mean
number of counts within 100 seconds was 100 $\pm$ 15 (mean $\pm$
standard deviation) i.e. about 1 count per second (which
corresponded to an accumulated equivalent dose of 0.040 $\pm$ 0.009
$\mu$Sv (mean $\pm$ standard deviation)  in 100
seconds)\footnote{According to the manufacturer~\cite{wedholm}, the
pulses can be collected from the neutron monitor by a BNC connector
which collects the pulses before the micro-computer which calculates
the equivalent dose and equivalent dose rate. The equivalent dose
and dose rate are collected instead from and RS232 output after the
micro-computer.}. These two pieces of information allowed us to
control that the BF$_3$ monitor was working in accordance with the
manufacturer specifications. A dose of 0.040 $\pm$ 0.009 $\mu$Sv in
100 seconds corresponds to an equivalent dose rate of 1.44 $\pm$
0.32 $\mu$Sv/h which is conformal to the dose rate of 1.5 $\pm$ 0.2
$\mu$Sv/h declared for the shielded source. Besides, being the mean
number of counts per second equal to 1 count/sec, the neutron
sensitivity coefficient is given by dividing it by the corresponding
dose rate 1.44 $\mu$Sv/h. Then, the neutron sensitivity of the
BF$_3$ is 0.7 cps/($\mu$Sv/h) which is conformal to what is declared
by the manufacturer (0.35 - 0.5 cps/($\mu$Sv/h) see~\cite{wedholm}).
Once that the correct operation and calibration of the neutron
monitor were established, we carried out two kinds of measurements
in order to determine the flux in cps/cm$^2$ (counts per second per
square centimetres) corresponding to the electronic noise of the
whole measuring apparatus and then the flux corresponding to the
background of the lab where the actual measurements had to be
carried out. The measurements to establish the electronic noise were
carried out in a shielded empty bunker of high density concrete
where the whole electronic equipment was brought and where the
BF$_3$ was surrounded over 4$\pi$ steradians by a further shield of
polyethylene blocks. Several 90 minute long measurements (the same
interval of the actual measurement) were carried out (sample size
$\geq$ 30) and the mean number of counts during 5400 seconds (90
minutes) was 20 $\pm$ 5 (mean $\pm$ standard deviation). We decided
to use the pessimistic value of 25 and hence we got 0.005 cps for
the electronic noise of the whole apparatus which corresponds to a
value of 0.03 $\cdot$ 10$^{-3}$ counts/(sec cm$^2$). Then we
measured the laboratory background flux. The measurements were
carried out with the same pulse collecting procedure that was going
to be used in the actual measurements. We carried out ten 180 minute
(10800 seconds) long measurements during which we registered the
number of pulses every 5 minutes (300 seconds) in order to get an
idea of the fluctuations of the flux during 180 minutes. The flux
turned out to be fairly constant around a mean value of 1.5 counts
within 300 seconds (i.e. 0.03 $\cdot$ 10$^{-3}$ counts/(sec~cm$^2$))
with some maximum fluctuations equal to 4.2 counts within 300
seconds (i.e. 0.1 $\cdot$ 10$^{-3}$ counts/(sec~cm$^2$)). In order
to put ourselves in the most pessimistic conditions we decided to
assume this maximum value as the laboratory background flux. The
electronic noise value 0.03 $\cdot$ 10$^{-3}$ counts/(sec cm$^2$)
and the background flux 0.1 $\cdot$ 10$^{-3}$ counts/(sec cm$^2$)
were summed and the result 0.13 $\cdot$ 10$^{-3}$ counts/(sec
cm$^2$) was adopted as the error to be attributed to each measured
value. The experimental set-up lay-out is schematically presented in
Fig.\ref{layout}. The BF$_3$ detector is placed next to the
cavitation chamber and its position with respect to the zone of
cavitation where the neutrons are expected to be emitted was the
same it had with respect to the AmBe source during the initial
calibration tests. The gamma meter UMo 1236 was next to the bottle
diametrically on the other side with respect to the BF$_3$. In
Fig.\ref{graph} we present an example of the neutron measurements
that were carried out during cavitation, i.e. the application of
ultrasounds to the 1000 ppm solution of Iron Chloride. As we already
said, the pulse collecting procedure was identical to that used
while measuring the background flux i.e. register the number of
pulses accumulated within 5 minutes.

Let's briefly describe the results depicted in Fig.\ref{graph}. The
graphic is divided into two parts by a vertical line (at 92 minutes
instead of 90 minutes for mere visual convenience) . The left side
from 0 to 90 minutes is the interval of time during which cavitation
was on. The right side from 90 to 180 minutes is the interval of
time during which cavitation was off but the neutron measurement
went on. In both sides, some peaks stand well above the background
level, pointing out that the emission of neutrons is not constant in
time, but occurs in bursts of neutrons or better in pulses. In the
left side of Fig.\ref{graph}, the first neutron pulse occurs after
40 minutes from the beginning of the cavitation and this
circumstance was the same for all the cavitation runs that were
carried out. More precisely, it turned out that in all cavitation
runs the first neutron pulse appeared 40 - 50 minutes after
switching on the ultrasounds . As to the right side of
Fig.\ref{graph}, although the cavitation was turned off and hence
one would expect that the neutron pulses would stop along with it,
there are two more peaks well above the background level. These
pulses, which were emitted in all cavitation runs after about 20
minutes the cavitation had been stopped, are a hysteretic behaviour
and a possible candidate explanation to this fact is that some of
the piezonuclear acoustic neutrons, (those neutrons emitted during
cavitation) had been absorbed most likely by the Carbon contained
both in the steel sonotrode and in the materials of the supporting
platforms and released after a latency (of about 20 minutes in our
case) as it normally occurs in the graphite of nuclear reactors.
Before drawing some conclusions we would like to point out three
details that, according to us, are important in order to create as a
clear and thorough picture of this new phenomenon as possible. As is
well known, it exists a phenomenon called fracto-emission or
fracto-fusion~\cite{kaushik,storms} during which neutron bursts can
be detected. Despite the fact that in most cases the cracked solids
are loaded with deuterium which is not our case, that our results
have nothing to do with fracto-emission has been proved by the
fairly large number of cavitation runs that we carried out under
different conditions. More precisely, we cavitated Aluminium
Chloride and Lithium Chloride solutions applying to them the same
ultrasonic power i.e. the same oscillation amplitude of the
sonotrode for the same amount of time and we did not register any
neutron peak like those in Fig.\ref{graph} but the level of the flux
was always compatible with the background. Besides, we performed
different cavitation runs of the same kind of solution both without
changing the sonotrode and by a new sonotrode for each run in order
to test possible effects due to the sonotrode aging and damaging. No
correlation of aging and damaging with neutron emission was found,
allowing us to exclude any possible implication of fracto-emission.
The second experimental fact that we want to highlight is related to
the association between cavitation and microscopical thermonuclear
fusion occurring at the bubble
collapse~\cite{tale1}-\cite{tale5},\cite{storms}. As we have already
stated, the solutions that we used and from which we obtained
neutrons did not contain any Deuterium, but only deionised
bidistilled water with some Iron Chloride. Besides, it seems
impossible to think of H$_2$-H$_2$  fusion since the two solutions
of Aluminium Chloride and Lithium Chloride and the simple deionised
bidistilled water did not produce any sign of neutrons above the
background level, although the first two had the same concentration
of that with Iron and all of them were treated with the same
ultrasonic power for the same amount of time. As we stated above
ionising radiation detection was carried out in parallel with
neutron detection. In particular gamma rays were monitored by the
UMo 1236 detector\footnote{The UMo1236 was calibrated by a Cobalt-60
standard source.}. The neutron and gamma monitoring proceeded
constantly in parallel for 180 minutes. The gamma response was in
equivalent dose rate and equivalent dose. The variations of the
gamma equivalent dose rate over 180 minutes were compared with the
variations of the cps/cm$^2$ from the neutron detector and
particular attention was given in order to make out possible gamma
peaks around those times when a neutron pulse occurred. Neither
coincidence nor correlation was found between neutron pulses and
gamma equivalent dose rate and dose which were always compatible
with the gamma background whose variations had been extensively
studied all over the lab and were of the order of 0.14 $\pm$ 0.05
$\mu$Sv/h (mean $\pm$ standard deviation) for equivalent dose rate
and 0.22 $\pm$ 0.07 $\mu$Sv (mean $\pm$ standard deviation) for
equivalent dose. Eventually we briefly want to say that a careful
statistical analysis of the pulse values collected in the different
cavitation runs was made in order to verify that they were not
normally distributed as it can be inferred from the theoretical
framework~\cite{carmig2,carmig3} that explains these new physical
phenomena (the graphical method called normal probability plot was
used for assessing whether or not the pulse values were normally
distributed). The values of the flux of the neutron pulses showed in
no way to have a gaussian distribution. Let's try and put forward
briefly some potential phenomenological thoughts about this lack of
gaussian distribution and the microscopical mechanism that steers
the neutron pulses. It has been said that the flux values of the
neutron pulses collected are not normally distributed as it was
ascertained by a normal probability plot test. This evidence tells
us that these values cannot be considered samples of a stochastic
variable with a gaussian distribution, since such a variable is
affected by independent small effects whose additive contributions
build the gaussian distribution. Conversely, this circumstance
indicates that there are some still unknown and above all
uncontrolled systematic causes that steer the microscopical
mechanism (the collapse of bubbles) that produces these neutron
pulses (like bubble dimensions, species and number of atoms trapped
on the bubble surface  etc.). The investigation of these causes is
the purpose of our present and future experiments. With this in
mind, we can look at the whole phenomenon of neutron pulse emission
by non radioactive nuclides subjected to ultrasounds from a
practical perspective and wherefrom put forward an analogy about its
behaviour. It is in fact possible to imagine this phenomenon to
behave like a neutron pulse generator randomly driven by a
multi-filter chopper able to change the frequency of the pulses,
their time duration and their intensity i.e. their flux, and the
energy of the neutrons emitted (a bit like the inverse of the famous
"neutron wheel made of paraffin used by Fermi to select neutrons of
different speed). In other words, we can look at this phenomenon as
a multi-varied neutron pulse source.

\section{Conclusions}

Mechanical vibrations in the form of ultrasonic oscillations of
suitable power and frequency were conveyed during 90 minutes by a
steel sonotrode into a solutions made of deionised bidistilled
water, Iron and Chlorine contained in a Boron Silicate glass vessel
of 250 ml. The rest of the stands and racks that were used to build
the experimental set-up were made of Iron, Aluminium,
Polychloroprene (Neoprene), Polymethyl Methacrylate (Plexiglas) that
being sold for the most different applications are not made of
unstable radioactive or fissile elements or even less contain large
quantities of light elements like Deuterium. Of course it goes
without saying that all the neutron and gamma laboratory background
measurements were intended both to fix a level of blindness for the
values that we were going to collect and also to double check the
remote chance of any sort of tiny activity of the above materials.
With this in mind, we state that the outcomes of the measurements
presented in this letter show for the first time that mechanical
vibrations can force absolutely stable nuclides (Iron is the most
stable in terms of binding energy per nucleon) to emit nuclear
radiation like neutrons and hence to undergo some sort of nuclear
reactions. The presence of neutrons from stable nuclides, their
emission in pulses, the timing of the pulses, the absence of gamma
radiation and the lack of normal distribution of the values of the
flux (cps/cm$^2$) corroborate the theoretical framework
~\cite{carmig1,carmig2,carmig3} on which the idea and the design of
these experiments was based. Let us add that these neutrons were
named piezonuclear neutrons because the phenomenon presented in this
letter is brought about by mechanical pressure (piezo from the
ancient Greek piezein which means to press) which affects the
nucleus of stable elements and makes it emit neutrons. Let us
conclude by putting forward a conjecture  about these piezonuclear
reactions and foretell that they can be brought about by properly
compressing solid materials that contain iron (e.g. granite), for
instance in one of those toughness experiments that are very common
in Mechanical and Civil Engineerings. More precisely, it will be
possible to measure neutron emission at the instant of fracture of
the specimens of these materials as their compression increases and
reaches the breaking load. According to what is being done for
liquids, it will be necessary to study neutron emissions as function
of the compression speed of the specimens.

\end{document}